\documentstyle[preprint,aps]{revtex}
\tightenlines
\begin{document}
\draft
\title{The classical wormhole solution and wormhole wavefuction with a nonlinear Born-Infeld scalar
 field}
\author{H. Q. Lu\footnote{ E--mail:alberthq\_lu@hotmail.com}, L. M. Shen,
P.  Ji, G. F. Ji and N. J. Sun} 
\address{Department of Physics, Shanghai University, Shanghai, 200436,
People's Republic of China}
\maketitle
\begin{abstract}
 On this paper we consider the classical wormhole
solution of the Born-Infeld scalar field. The corresponding classical
wormhole solution can be obtained analytically for both very small and large 
$\dot{\varphi}$. At the extreme limits of small $\dot{
\varphi}$ the wormhole solution has the same format as one
obtained by Giddings and Strominger[10]. At the
extreme limits of large $\dot{\varphi}$ the wormhole
solution is a new one. The wormhole wavefunctions can also be obtained for
both very small and large $\dot{\varphi}$. These wormhole
wavefunctions are regarded as solutions of quantum-mechanical
Wheeler--Dewitt equation with certain boundary conditions.
\end{abstract}


\section{Introduction}

The corresponding Lagrangian of Born-Infeld field has been first
proposed by Heisenberg[3] in order to describe the process of
meson multiple production connected with strong field regime, as a
generalization of the Born-Infeld one, $L^{BJ}=b^2\left[ \sqrt{1-\left(
1/2b^2\right) F_{ik}F^{ik}}-1\right]$[4], that removes the
point--charge singularity that means classical electrodynamics. When the
parameter of the field approaches to zero, the corresponding Lagrangian will
reduce to linear case[3,4]. Born-Infeld type Lagrangians have
also been considered in the theory of strings and branes as well as gravity
[5-9]. It shows that the low energy effective field theory on
D-branes is of Born-Infeld type[5]. The consistency of the 
$\sigma $--model for the world sheet of string is shown to require that the
brane should be described by Born-Infeld action, just like in the general
curved background requiring consistency of string theory leads to the
Einstein-Hilbert action.

According to the Euler-Lagrangian equation of motion of Born-Infeld scalar
field , we can obtain $\dot{\varphi}$ at the limit of large and small
cosmological scale factors $R$ respectively. At such limit condition, we
found classical wormhole and quantum wormhole solutions. This paper
comprises following contents: In section 2 we obtain the classical wormhole
solution of Born-Infeld scalar field. In section 3 we found wormhole
wavefunction of our nonlinear scalar field model. In last section, we
discuss our results and come to conclusions.

\section{Classical wormhole solution}

The Euclidean action of gravitational field interacting with a Born-Infeld
type scalar field is given by 
\begin{equation}
S_E=\int \frac{R_c}{16\pi G}\sqrt{g}d^4x+\int L_s\sqrt{g}d^4x  
\end{equation}
Where\ we have chosen unit so that $c=1$, $R_c$ is the Ricci scalar
curvature and the Lagrangian $L_s$ of the nonlinear Born-Infeld scalar field
is [3] 
\begin{equation}
L_s=\frac 1\lambda \left[ 1-\sqrt{1-\lambda \varphi ,_{\mu }\varphi
,_{\nu }g^{\mu \nu }}\right] 
\end{equation}
When $\lambda \longrightarrow 0$, based on Taylor expansion (2)
approximates to 
\begin{equation}
\lim_{\lambda\rightarrow 0}L_s=\frac 12\varphi ,_{\mu
}\varphi ,_{\nu }g^{\mu \nu } 
\end{equation}
We choose the standard Euclidean and closed R-W metric 
\begin{equation}
ds^2=d\tau ^2+R^2\left( \tau \right) \left\{ \frac{dr^2}{1-r^2}+r^2\left[
\left( d\theta ^2\right) +\sin ^2\theta \left( d\varphi ^2\right) \right]
\right\}  
\end{equation}
Where $\tau $ is the Euclidean radial coordinate and $R\left( \tau \right) $
is the radius of curvature of a 3D sphere. According to the ``cosmological
principle'', $R$ must only depend on $\tau $. We write Einstein equations as 
\begin{equation}
-\frac{3\dot{R}^2}{R^2}+\frac 3{R^2}=8\pi GT_{\;\;0}^0  
\end{equation}
\begin{equation}
-\frac{2\ddot{R}}R-\frac{\dot{R}}{R^2}+\frac 1{R^2}=8\pi
GT_{\;\;1}^1=8\pi GT^{2}_{\;\;2}=8\pi GT_{\;\;3}^3  
\end{equation}
Where the upper index ``$\cdot $'' denotes the derivative with respect to
$\tau $. We substitute the Lagrangian (2) into Eule-Lagrange equation 
\begin{equation}
\frac d{d\tau }\left( \frac{\partial L_s}{\partial \dot{\varphi}}\right) -
\frac{\partial L_s}{\partial \varphi }=0  \end{equation}
Then we obtain 
\begin{equation}
\frac{R^6\dot{\varphi}^2}{1+\lambda \dot{\varphi}^2}=W_0  
\end{equation}
and consequently 
\begin{equation}
\dot{\varphi}=\sqrt{\frac{W_0}{R^6-W_0\lambda }}  \end{equation}
Where $W_0$ is a constant of integration. We write components of
energy-momentum tensor of Born-Infeld scalar field as 
\begin{equation}
T_{\;\;\nu }^{^\mu }=\frac{g^{\mu \rho }\varphi ,_{\rho }\varphi ,_{\nu }}{
\sqrt{1-\lambda \varphi ,_{\mu }\varphi ,_{\nu }g^{\mu \nu }}}-\delta
_{\nu }^{\mu }L_s  \end{equation}
Substitute equations (9) and (2) into (10), we obtain 
\begin{equation}
T_{\;\;0}^0=\frac 1\lambda \left[ \sqrt{R^6-\lambda W_0}/R^3-1\right] 
\end{equation}

\begin{equation}
T_{\;\;1}^1=T_{\;\;2}^2=T_{\;\;3}^3=-\frac 1\lambda \left[ 1-R^3/\sqrt{R^6-\lambda
W_0}\right]  \end{equation}
Substitute (11) into Einstein equations (5), we can obtain 
\begin{equation}
-\frac{3\dot{R}^2}{R^2}+\frac 3{R^2}=\frac{8\pi G}\lambda \left[ \frac{
\sqrt{R^6-\lambda W_0}}{R^3}-1\right]  
\end{equation}

\begin{equation}
\dot{R}^2=1-\frac{8\pi G}{3\lambda }\left[ R^2\sqrt{1-\lambda W_0R^{-6}}
-R^2\right]  
\end{equation}
From equation (9), we can find that $R$ is very small or very large when $\dot{%
\varphi}$ is very large or very small respectively.
Assuming that $\dot{\varphi}$ is very small (i.e. $R$ is very large), equation
(14) becomes
\begin{equation}
\dot{R}^2=1+\frac{4\pi GW_0}{3R^4}  \end{equation}
When $W_0<0$, the wormhole solution of equation (15) is 
\begin{equation}
\frac \tau {R_0}=\sqrt{\frac 12}F\left[ \cos ^{-1}\left( \frac{R_0}
R\right) ,\sqrt{\frac 12}\right] -\sqrt{2}E\left[ \cos ^{-1}\left( \frac{
R_0}R\right) ,\sqrt{\frac 12}\right] +\frac{\sqrt{R^4-R_0^4}}{R_0R} 
\end{equation}
Where $R_0=\sqrt[4]{\frac{-4\pi GW_0}3}$.
We note that this wormhole solution has the same format as one obtained by
Giddings and Strominger[10]. When $\dot{\varphi}$ is very large
(i. e., $R$ is very small), we can obtain from equation (14) 
\begin{equation}
\dot{R}^2=1-\frac{8\pi G\sqrt{\frac{-W_0}{9\lambda }}}R  
\end{equation}
We restrict $\lambda >0$, integrating (17) we can obtain wormhole solution
of equation (17), that is 
\begin{equation}
R\sqrt{1-\frac NR}+N\log \left[ \frac{\sqrt{\frac RN-1}+\sqrt{\frac RN}-1}
{\sqrt{\frac RN-1}-\sqrt{\frac RN}+1}\right] =\tau  \end{equation}
Where $N=8\pi G\sqrt{\frac{-W_0}{9\lambda }}>0$.
From equation (18) we can find that $\lim_{\tau \rightarrow \infty } 
R\left( \tau \right) =\infty $. Using $\dot{R}\left( 0\right) =0$ from equation
(17) we can obtain the size of wormhole throat: $R\left( 0\right) =N$ and 
$\ddot{R}\left( 0\right) =\frac 1{2N}$. Thus we obtain a new wormhole
solution.

\section{Wormhole wavefunction}

It is possible that the wormholes are regarded as solutions of
quantum-mechanical Wheeler ---Dewitt(WD) equation. These wavefunctions have
to obey certain boundary conditions in order that they represent wormholes.
The wavefunction will be damped at large radius $R$, i.e., such wavefunction
tends to zero as $R\rightarrow \infty $, and when $R$ nears $0$ it should be
oscillatory[11]. Wavefunction should tend to a constant as 
$R\rightarrow 0$[12]. The Lorentz action of the gravitational
field interacting with a Born--Infeld type scalar field is given by 
\begin{equation}
S=\int \frac{R_c}{16\pi G}\sqrt{-g}d^4x+\int L_s\sqrt{-g}d^4x  
\end{equation}
Where $R_c$ is the Ricci scalar curvature and $L_s$ is equation (2). However,
in equation (19) , $g_{\mu \nu }$ is decided by equation (20).The closed R--W
spacetime metric is 
\begin{equation}
ds^2=-dt^2+R^2\left( t\right) \left\{ \frac{dr^2}{1-r^2}+r^2\left[ d\theta
^2+\sin ^2\theta \left( d\varphi ^2\right) \right] \right\}  
\end{equation}
Using equation (20) and integrating space--components, the action (19) becomes
(The upper-dot means the derivative with respect to the time $t$.): 
\begin{equation}
S=\int \frac{3\pi }{4G}\left( 1-\dot{R}^2\right) Rdt+\int 2\pi ^2R^3\left[
\frac 1\lambda \left( 1-\sqrt{1-\lambda \varphi ,_{\mu }\varphi ,_{\nu
}g^{\mu \nu }}\right) \right] dt\equiv \int {\cal L}_gdt+\int {\cal L}_s
dt  \end{equation}
To quantize the model, we first find the canonical moment $P_R=\left( \frac{
\partial {\cal L}_g}{\partial \dot{R}}\right) =-\left( \frac{3\pi }{2G}
\right) \smallskip R\dot{R}$ , $P_{\varphi }=\left( \frac{\partial 
{\cal L}_s}{\partial \dot{\varphi}}\right) =\left( 2\pi ^2R^3\dot{\varphi}
/\sqrt{1+\lambda \dot{\varphi}^2}\right) $ and the Hamiltonian $H=P_R\dot{R}
+P_{\varphi }\dot{\varphi}-{\cal L}_g-{\cal L}_s$, 
\begin{equation}
H=-\frac G{3\pi R}P_R^2-\frac{3\pi }{4G}R+\frac{2\pi ^2R^3}\lambda \left(
1-\sqrt{1-\frac{\lambda P_{\varphi }^2}{4\pi ^4R^6}}\right)  \end{equation}
For small $\dot{\varphi}$, the Hamiltonian (22) can be simplified by using
the Taylor expansion 
\begin{equation}
H=-\frac G{3\pi R}P_R^2-\frac{3\pi }{4G}R+\frac{P_{\varphi }^2}{4\pi
^2R^3}-\frac{\lambda P_{\varphi }^4}{64\pi ^6R^9}  \end{equation}
If $\dot{\varphi}$ is large, then equation (22) becomes 
\begin{equation}
H=-\frac G{3\pi R}P_{R}^2-\frac{3\pi }{4G}R+\frac{2\pi ^2R^3}\lambda 
\end{equation}
The WD equation is obtained from $\hat{H}\psi =0$ and equations (23) as well as
(24) by replacing $P_R\rightarrow -i\left( \frac \partial {\partial
R}\right) $ and $P_{\varphi }\rightarrow -i\left( \frac \partial {\partial
\varphi }\right) $. Then we obtain 
\begin{equation}
\left[ \frac{\partial ^2}{\partial R^2}+\frac PR\frac \partial {\partial
R}-\frac 1{R^2}\frac \partial {\partial \Phi ^2}-\frac \lambda {16\pi
^4R^8}\frac{\partial ^4}{\partial \Phi ^4}-U\left( R\right) \right] \psi =0
\end{equation}
and 
\begin{equation}
\left[ \frac{\partial ^2}{\partial R^2}+\frac PR\frac \partial {\partial
R}-u\left( R\right) \right] \psi =0  \end{equation}
Where $\Phi ^2=4\pi G\varphi ^2/3$ and the parameter $P$ represents the
ambiguity in the ordering of factors $R$ and $\frac \partial {\partial R}$
in the first term of equations (23) and (24). We have also denoted
$$U\left( R\right) =\left( \frac{3\pi }{2G}\right) ^2R^2$$ 
$$u\left( R\right) =\left( \frac{3\pi }{2G}\right) ^2R^2\left[ 1-\frac{8\pi G}
{3\lambda }R^2\right]$$ 
Equations (25) and (26) are the WD equations corresponding to action (19) in
the cases of small and large $\dot{\varphi}$ respectively. Together we can
obtain the equation of motion of Born--Infeld scalar field when we
substitute the Lagrangian $L_s$ into the Eule--Lagrangian equation 
\begin{equation}
\frac d{dt}\left( \frac{\partial L_s}{\partial \dot{\varphi}}\right) -
\frac{\partial L_s}{\partial \varphi }=0  
\end{equation}
Then we obtain 
\begin{equation}
\dot{\varphi}=\sqrt{\frac C{R^6+C\lambda }}  
\end{equation}
The upper--dot means the derivative with respect to the $t$. Where $C$ is a
constant of integration. From equation (28) we find that $R$ is very small or
very large when $\dot{\varphi}$ is very large or very small respectively.
In other word, equations (25) and (26) are the WD equations corresponding to
action (19) in the cases of large and small $R$ respectively.
When $R$ is very large, we take the ambiguity of ordering factor $P=-1$ and
set transformation $\left( R/R_0\right) ^2=\sigma $, with $R_0$ the Planck's
length. Choosing appropriate units makes the Planck constant $\hbar =1$,
speed of light $c=1$, and $R_0\sim \sqrt{\frac{4G}{3\pi }}$. Then equation (25)
becomes
\begin{equation}
\frac{\partial ^2\psi }{\partial \sigma ^2}-\frac 1{\sigma ^2}\frac{
\partial ^2\psi }{\partial \Phi ^2}-\frac \lambda {16\pi ^4R_0^6\sigma ^5}
\frac{\partial ^4\psi }{\partial \Phi ^4}-\stackrel{\sim }{U}\psi =0 
\end{equation}
Where $\stackrel{\sim }{U}=\left( 3\pi /4G\right) ^2R_0^4$. Assuming $\psi
\left( \sigma ,\Phi \right) =Q\left( \sigma \right) e^{-K\Phi }$, with $K$
an arbitrary constant, equation (29) takes the form: 
\begin{equation}
\frac{d^2Q}{d\sigma ^2}-\left( \frac{K^2}{\sigma ^2}+\frac{\mu K^4}{
\sigma ^5}+\stackrel{\sim }{U}\right) Q=0  
\end{equation}
Where $\mu =\lambda /16\pi ^4R_0^6$. When $R$ (and consequently $\sigma $) is
very large, equation (30) approximates to 
\begin{equation}
\frac{d^2Q}{d\sigma ^2}-\beta ^2Q=0  \end{equation}
Where $\beta =\left( \frac{3\pi }{4G}\right) R_0^2$ .
The solution of equation (31) is 
\begin{equation}
Q=\exp \left( -\beta \sigma \right) \end{equation}
From (32) we can find that wavefunction $\psi \rightarrow 0$ when $%
R\rightarrow \infty $ (and consequently $\sigma \rightarrow \infty $).
If $R$ is very small, we take the ambiguity of ordering factor $P=-1$ and
set the transformation $\left( R/R_0\right) ^2=\sigma $ ,with $R_0$ the
Planck length. Choosing appropriate units makes the Planck constant $\hbar
=1$, the speed of light $c=1$ and $R_0\sim \sqrt{\frac{4G}{3\pi }}$.
Then equation (26) becomes 
\begin{equation}
\frac{d^2\psi }{d\sigma ^2}-\left( \frac{3\pi }{4G}\right) ^2R_0^4\left( 1-
\frac{8\pi G}\lambda \sigma R_0^2\right) \psi =0  \end{equation}
When $R>>\sqrt{\frac{3\lambda }{8\pi G}}$(and consequently $\sigma >>\frac{
3\lambda }{8\pi GR_0^2}$), equation (33) can be approximated as 
\begin{equation}
\frac{d^2\psi }{d\sigma ^2}+\gamma ^2\sigma \psi =0  \end{equation}
Where $\gamma =\left( \frac{3\pi ^3}{2G\lambda }\right) ^{1/2}R_0^2$. Equation
(34) has the solution 
\begin{equation}
\psi =\sqrt{\sigma }Z_{\frac 13}\left( \frac{2\gamma }3\sigma ^{3/2}\right)
\end{equation}
Now $\psi $ is an oscillatory function. When $\sigma \rightarrow 0$ (and
consequently $R\rightarrow 0$), equation (33) can be approximated as 
\begin{equation}
\frac{d^2\psi }{d\sigma ^2}-\left( \frac{3\pi }{4G}\right) ^2R_0^4\psi =0 
\end{equation}
Equation (36) has the solution 
\begin{equation}
\psi =Ne^{-\frac{3\pi }{4G}R^2}  \end{equation}
In the geometry described by the R-W metric, the probability of wormhole
situated between $R\rightarrow R+dR$ is 
\begin{equation}
\omega \left( R\right) \propto \psi ^2R^2dR  \end{equation}
The probability density is $\psi ^2R^2$. The position of the maximum
probability can be determined by 
\begin{equation}
\frac d{dR}\left( \psi ^2R^2\right) =0  \end{equation}
From (39) we can obtain 
\begin{equation}
R=\sqrt{\frac{4G}{6\pi }}  
\end{equation}
Equation (40) implies that most probable radius of wormhole is of the Planck
scale, namely the quantum effect can make a wormhole survive gravitational
collapse.

\section{Conclusion}

At the extreme limits of small $\dot{\varphi}$, the classical wormhole
solution of the Born--Infeld scalar field has the same format as one
obtained by Giddigs and Strominger. If $\dot{\varphi}$ is very large, a new
wormhole solution can be obtained.
From the Eule--Lagrange equation of the Born--Infeld scalar field, we find
that cosmological scale factors is very large or very small when $\dot{
\varphi}$ is very small or very large respectively. We obtain the wormhole
wavefunction. It is the solution of quantum--mechanical Wheeler--Dewitt
equation with certain boundary conditions. The wavefunction is exponentially
damped for large three geometries and the wavefunction tends to a zero as
cosmological scale factors tend to a infinity. They oscillate near zero
radius, it tends to a constant as cosmological radius tends to a zero.

\acknowledgments 
Part of this work is supported by the National Natural Science Foundation of
China under grant No. 10073006 and the National Science Foundation of
Shanghai under grant No. 00zd14018.


\begin{references}

\bibitem{[1]} Hawking S W 1987 {\it Phys. Lett.} {\bf 195B} 337

\bibitem{[2]} Hawking S W 1988 {\it Phys. Rev.} D {\bf 37} 904

\bibitem{[3]} Heisenberg W 1952 Z. {\it Phys.} {\bf 133} 79

\bibitem{[4]} Born M and Infeld Z 1934 {\it Proc. Roy. Soc.} A {\bf 144} 425

\bibitem{[5]} Tseytlin A 1986 {\it Nucl. Phys.} B {\bf 276} 391

\bibitem{[6]} Palatnik D 1998 {\it Phys. Lett.} {\bf 432B} 287

\bibitem{[7]}  Feigenbaum J A 1998 {\it Phys. Rev.} D {\bf 58} 124023

\bibitem{[8]} Boillat G and Strumia A 1998 {\it J Math. Phys.} {\bf 40} 1

\bibitem{[9]} Deser S and Gibbons G W 1998 {\it Class. Quantum Grav.} {\bf 15} 135

\bibitem{[10]} Giddings S B and Strominger A 1988 {\it Nucl. Phys.} B {\bf 306} 890

\bibitem{[11]} Hawking S W and Page D N 1990 {\it Phys Rev.} D {\bf 42} 2655

\bibitem{[12]} Coule D H 1992 {\it Class. Quantum Grav.} {\bf 9} 2353-2360
\end{references}
\end{document}